\documentclass[%
 aip,
 sd,%
 amsmath,amssymb,
reprint,%
]{revtex4-1}

\usepackage{graphicx}
\usepackage{dcolumn}
\usepackage{bm}
\usepackage{longtable}
\usepackage{epsfig}
\usepackage{color}
\usepackage{setspace}

\begin{document}

\preprint{JCP/001-IrO}

\title[]{Relativistic configuration interaction calculation on the ground and excited states of iridium monoxide}

\author{Bingbing Suo}
\affiliation{Institute of Modern Physics, Northwest University, Xi'an, Shaanxi 710069, China}
\author{Yan-Mei Yu}
\email{ymyu@aphy.iphy.ac.cn}
\affiliation{Beijing National Laboratory for Condensed Matter Physics, Institute of Physics, Chinese Academy of Sciences, P.O.Box 603, Beijing 100190, China}
\author{Huixian Han}
\email{hxhan@nwu.edu.cn}
\affiliation{School of Physics, Northwest University, Xi'an, Shaanxi 710069, China}

\date{\today}

\begin{abstract}
We present the fully relativistic multi-reference configuration interaction calculations of the ground and low-lying excited electronic states of IrO for individual spin-orbit component. The lowest states for four spin-orbit components 1/2, 3/2, 5/2, and 7/2 are calculated intensively to clarify the ground state of IrO. Our calculation suggests that the ground state is of 1/2 spin-orbit component, which is highly mixed with $^4\Sigma^-$ and $^2\Pi$ states in $\Lambda-S$ notation. The two low-lying states of the 5/2 and 7/2 spin-orbit components are nearly degenerate with the ground state and locate only 234 and 260 cm$^{-1}$ above, respectively. The equilibrium bond length 1.712 \AA \  and harmonic vibrational frequency 903 cm$^{-1}$ of the 5/2 spin-orbit component are close to the experimental measurement of 1.724 \AA \  and 909 cm$^{-1}$, which suggests the 5/2 state should be the low-lying state contributed to spectra in experimental study. Moreover, the electronic states that give rise to the observed transition bands are assigned in terms of the excited energies and oscillator strengths obtained for the 5/2 and 7/2 spin-orbit components.
\end{abstract}

\pacs{31.15.Vn, 31.15.ae, 31.15.ag}
\keywords{Iridium monoxide, electronic state, spectrum, relativistic MRCI}
\maketitle

%

\section{Introduction}

Though Iridium Monoxide is simple diatomic molecule, it is a difficult system both on experimental and theoretical studies due to the existence of nearly degenerated open 5d shells and also the notably relativistic effect of the Ir atom. The multiconfigurational character of electron structure of IrO could give rise to strongly perturbed electronic transitions, which would produce complicated spectra in the experimental study. In particular, as a heavy-element with open d shell, Ir is an typically interesting system that requires accurate description of electron correlation, both static and dynamic, and also the relativistic effect contributions in theoretical study. This poses a great challenge to the modern electron structure theory.

The emission spectra of IrO was first recorded by Raziunas \textit{et al.}, who observed four bands situated at 5856, 5990, 6972 and 6899{\AA} \cite{Raziunas-JCP-1965}, respectively, but no rotational and vibrational analysis is made because of low resolution of the spectra. In 1972, Jansson and Scullman \cite{Jansson-JMS-1972} analyzed fifteen emission bands in the region of 4200-6400 {\AA}, and one of them is assigned as the band of 5950 {\AA} observed in Raziunas's study \cite{Raziunas-JCP-1965}. Three subsystems with $\Delta \Omega=0$ are resolved and assigned as sub-state transitions between two $^2\Delta$ states \cite{Jansson-JMS-1972}. After that, several theoretical studies have been carried out for the IrO molecule \cite{Citra-JPCA-1999,Song-TCA-2007,Suo-JPB-2012}. The first theoretical work is performed by Citra and Andrews on density function theory (DFT) \cite{Citra-JPCA-1999}. They assigned $^4\Sigma^-$ as the ground state of IrO. Later, a systemic study of 5d metal oxide by Yao \textit{et al.} also on DFT level of theory supported this conclusion \cite{Song-TCA-2007}. In general, DFT is insufficient to describe electron structure of such complicate molecule as IrO because the static correlation is difficult to be caught by a single determinate approach.

Therefore, two authors in present study, Suo and Han have studied 25 low-lying electronic states of IrO by using the multi-state complete active space second order perturbation theory (MS-CASPT2) method. Some important states are calculated by the restrict active space state interaction (RASSI) method to take into account the spin-orbit coupling (SOC) effect \cite{Suo-JPB-2012}. Different with previous theoretical prediction, the $^4\Delta_{7/2}$ is assigned as the ground state of IrO. If the spin-orbital coupling effect is omitted, the $^2\Pi$ state is lowest in energy. Almost at same time with Suo \textit{et al.}'s theoretical work, Pang, \textit{et al.} have recorded five electronic transition spectra of IrO in the region between 448 and 650 nm, which are identified as [17.6] 2.5-X$^2\Delta_{5/2}$, [17.8] 2.5-X$^2\Delta_{5/2}$, [21.5] 2.5-X$^2\Delta_{5/2}$, [22.0] 2.5-X$^2\Delta_{5/2}$, and [21.9] 3.5-3.5 systems \cite{Pang-JPCA-2010}. According to the observed transitions, they concluded that IrO has a X$^2\Delta_{5/2}$ ground state. Obviously, the ground state $^4\Delta_{7/2}$ obtained from high level theoretical calculation is inconsistent with Pang \textit{et al.}'s assignment of the $\Omega=5/2$ ground state. In Suo \textit{et al.}'s study, a state with $\Omega=5/2$ lies 2425 $cm^{-1}$ above the ground state and is dominated by $^4\Delta_{5/2}$. The $^2\Delta_{5/2}$ is higher in energy and locates at 8310 $cm^{-1}$ above the $^4\Delta_{7/2}$ (In Suo \textit{et al.}'s work \cite{Suo-JPB-2012}, the $^2\Delta_{5/2}$ is mislabeled as $^2\Delta_{3/2}$ and the $^2\Delta_{3/2}$ is mislabeled as $^2\Delta_{1/2}$). However, Pang \textit{et al.}'s experiment is performed on cold molecular beam, the energy separation 2425 $cm^{-1}$ between $^4\Delta_{7/2}$ and $^4\Delta_{5/2}$ is too large to be overcome by thermal fluctuation. Therefore, the low-lying state in the experimental spectra should almost be the ground state.

Inspired by the serious discrepancy between the theoretical and experimental studies, Adam \textit{et al.} have re-examined spectra of IrO via high resolution LiF experiments \cite{Adam-JMS-2013}. Two transitions $[17.5]2.5-X2.5$ and $[23.3]2.5-X2.5$ are fully resolved and the ground state is verified to have spin-orbit component $\Omega$ value of $5/2$ \cite{Adam-JMS-2013}. These researchers proposed that the $^2\Delta_{5/2}$ state may interact with $^4\Delta_{5/2}$ strongly and push the later below the $^4\Delta_{7/2}$ state \cite{Adam-JMS-2013}. According to Adam \textit{et al.}'s suggestion, Suo and Han include $A^2\Delta_{5/2}$ into SOC calculation and find that the first state with $\Omega=5/2$ is highly mixed with $^4\Delta_{5/2}$, $A^2\Delta_{5/2}$ and $B^2\Delta_{5/2}$ \cite{Suo-supplement}. The energy gap between the $\Omega=5/2$ and $\Omega=7/2$ is reduced to 1108 cm$^{-1}$. However, such energy gap is still too large to identify the $\Omega=5/2$ ground state doubtlessly.

With increase of the nuclear charge, the relativistic effect becomes crucial in theoretical description of the chemistry of heavy elements. Therefore, j-j coupling is more appropriate than L-S coupling for interpretation of the experimental spectra. However, Suo et al.'s calculation reckons the scalar relativistic effect via spin-free second order Douglas-Kroll Hamiltonian (DKH) Hamiltonian \cite{DKH1974,DKH1985,DKH1986}. Then, SOC was treated as the perturbation and evaluated via RASSI \cite{RASSI}. This calculation may not sufficient for IrO due to the strong spin-orbital coupling in heavy atom leaves only total angular moment J as a good quantum number. Therefore, in this work, the electron states of IrO are calculated for individual J components by using fully relativistic multi-reference configuration interaction method. The relativistic effects and the electronic correlations are considered on the same footing in Kramers-restricted configuration interaction (KRCI) implementation based on the exact 2-Component (X2C) Hamiltonian \cite{X2C_Liu, X2C_Ilias}. We present and discuss the electronic ground and excited states of $\Omega$=1/2, 3/2, 5/2, and 7/2 of IrO. Our calculated results show the strong degeneracy of the lowest state with $\Omega$=1/2, 5/2, and 7/2, which can explain the discrepancy between the calculated and observed ground states. In addition, the excited stated obtained for $\Omega$=5/2 and 7/2 are used to assign the observed six transition bands of LIF spectrum in the visible region.

\section{Computational method}
All our calculations utilize the exact two-component (X2c) Hamiltonian that includes by default atomic-mean-field two-electron spin-same-orbit corrections \cite{X2C_Liu, X2C_Ilias}. Firstly, average-of-configuration Dirac-Hatree-Fock (AC-DHF) \cite{Tyssen-DHF} or Kramers restricted multi-configuration self consistent field (KR-MCSCF) \cite{KR-MCSCF} are performed to optimize molecular spinors in which static correlation has been evaluated. In AC-DHF and KR-MCSCF calculations, 13 electrons are allowed to distribute in 9 Kramers pairs (13in9) consisting of the Ir 5d6s and O 2p orbitals. Then, the Kramers-restricted multi-reference configuration interaction (KR-MRCI) are carried out by using molecular spinors from AC-DHF or KR-MCSCF calculations \cite{KRCI2010}. It takes advantage of the concept of generalized active spaces (GAS) to define suitable correlation spaces thereby allowing for arbitrary occupation constraints. In our calculation, two sets of GAS are applied. The first GAS, namely, ``SD8(13in9)SD" includes the frozen core [$1s^2$ (O) and $1s^2 2s^2 2p^6 3s^2 3p^6 3d^{10} 4s^2 4p^6 4d^{10}4f^{14}5s^2$ (Ir)], the outer core [$2s^2$ (O) and $5p^6$ (Ir)], the valence [$2p^4$ (O) and $5d^7 6s^2$ (Ir)], and the virtual shells less than 2 a.u. The notation ``SD8(13in9)SD" means that at least two holes in the 8 electron of the outer core is allowed, 13 electrons are distributed in the valence shell with plus of excited electrons from the outer core to valence shells, and finally that all single and double excitations into virtual orbitals are taken into account. Therefore, the first GAS will bring 21 electrons at most into correlation. The second GAS ``SD20(15in10)SD" has a larger outer core [$4f^{14} 5p^6$ (Ir)], the valence [$2s^2 2p^4$ (O) and $5d^7 6s^2$ (Ir)], and the virtual orbitals less than 2 a.u, which brings 35 electrons at most into correlation. The quantum number for each individual electronic state is assigned through calculating the expectation value for the one-electron operator $j_z$=$l_z$+$s_z$. All our calculations are implemented with the relativistic quantum chemistry calculation package DIRAC \cite{DIRAC}.

In order to investigate the ground state of IrO, the four lowest-lying states for $\Omega$=1/2, 3/2, 5/2, and 7/2 are calculated by employing the Dunning's aug-cc-pCVTZ basis set \cite{Dunning} for O and Dyall's cv3z basis set for Ir \cite{Dyall}. For understanding the ground electronic structure of IrO, we also conduct the comparative calculation of CoO and RhO by using Dunning's aug-cc-pCVDZ for O \cite{Dunning} and dyall's cv2z basis sets for Co and Rh \cite{Dyall}. The excited states of IrO are calculated by the additional KRCI implementation. The adiabatic energy is obtained by using Dunning's aug-cc-pCVDZ (O) and dyall's cv2z basis sets (Ir), which is modified with the vertical excited energy calculated by using Dunning's aug-cc-pCVTZ (O) and dyall's cv3z basis sets (Ir) (See supplemental material for computational details). All calculations were performed using uncontracted basis sets. The expectation values $\langle s_z\rangle=\Sigma_{i}m_s$ and $\langle l_z\rangle=\Sigma_{i}m_l$ are given that suggest the major component in the $\Lambda-S$ notation. The KRCI prosperity module gives the transition dipole moment and therefore the oscillation strength is obtained in terms of
\begin{equation}
f=\frac{2m}{e^2\hbar\omega}|D|^2
\end{equation}
, where $\hbar\omega$ and $D$ is the excited energy and transition dipole moment, respectively, $e$ and $m$ is electronic charge and mass. The oscillation strength helps us to assign the excited states to the observed visible transition bands.

\section{Results and discussion}

\subsection{The ground state of IrO}

\begin{figure}
\begin{center}
  \includegraphics[width=8cm]{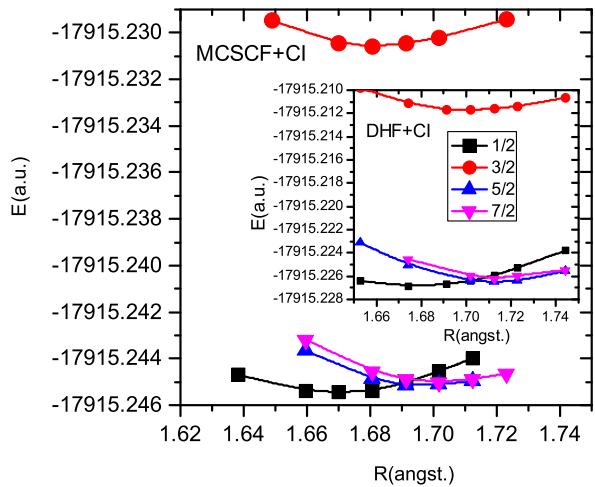}
\end{center}\vspace{-20mm}
\caption{PECS of the lowest states with $\Omega$=1/2, 3/2, 5/2, and 7/2, as calculated at the KRCI SD8(13in9)SD level implemented by the DHF+CI and MCSCF+CI ways.}
\label{fig1}
\end{figure}

We first compare the ground state of IrO on the different level of calculations. In Table \ref{tbl:groundstate}, we illustrate the equilibrium bond lengths, energy separations, harmonic vibrational frequencies and major configurations of each lowest $\Omega$ states on three different KRCI calculations. The first two calculations employ the same GAS SD8(13in9)SD but with different molecular spinors in which one is obtained by AC-DHF and the other is from KR-MCSCF. It is found that the two different implementations, the DHF+CI and MCSCF+CI, yield the almost same results for the spectroscopy constant and electronic configuration. This suggests that the AC-DHF has captured sufficient static correlation, even without more expensive KR-MCSCF calculations. Both of two calculations prefer an $\Omega=1/2$ ground state. The 5/2 and 7/2 states are slightly higher in energy and locates at 58 and 91 cm$^{-1}$ above the ground state on the MCSCF+CI calculation. The 3/2 state is well separated with three low-lying states and locates 2632 cm$^{-1}$ above the 1/2 state.

It is noteworthy that the energy separations of 1/2, 5/2 and 7/2 are so small that the energy gap of these three states may sensitive to different level of theories. Therefore, we have performed an extra KRCI calculation with a larger GAS space, in which the 4f shell is included in correlation calculation and is labeled as GAS SD20(15in10)SD. This calculation sufficiently considered correlations of the 4f electrons of Ir atom, and was our highest level of calculation in present work. As can be seen from Table \ref{tbl:groundstate}, the energy order of four lowest states does not change when the GAS space is enlarged from SD8(13in9)SD to SD20(15in10)SD. However, the energy gap of the 1/2 and 5/2 states is increased to 234 cm$^{-1}$, which is larger than the corresponding value 58 on SD8(13in9)SD calculation. Interestingly, the energy separation between the 5/2 and 7/2 states is still small with the value of 26 cm$^{-1}$, and is close to 23 cm$^{-1}$ obtained by the small-core calculation without the 4f electrons. Including 4f electrons in correlation calculation introduces more core-core and core-active correlations than small-core calculation. If assuming core-core correlation is similar in each states, it is correlation between the 4f and active electrons that stabilizes the 1/2 state and pushes this state lower in energy than SD8(13in9)SD result. The energy variation between the small-core and large-core calculations is minor, for example the energy separation of the 1/2 and 5/2 states is changed less than 200 cm$^{-1}$. Such small variation is not supposed change the order of the high lying excited state for a given $\Omega$ value. Therefore, we still use the small-core calculation in the following calculation for the excited states because of the economical computational cost. We intend to give a description of the excited states of IrO at a qualitative level that is sufficient for assignment of the experimental spectra.

\begin{spacing}{1}
\begin{table*}\small
\caption{The lowest-lying electronic states of IrO for $\Omega$=1/2, 3/2, 5/2, and 7/2, as calculated at the KRCI SD8(13in9)SD level in the DHF+CI and MCSCF+CI implementations and KRCI SD20(15in10)SD level in the MCSCF+CI, where $\langle s_z \rangle$ and $\langle l_z \rangle$ are expectation value of $s_z$ and $l_z$ operators, and $R$, $T$, $\omega$ are the equilibrium bond length ({\AA}), excited energy (cm$^{-1}$), and harmonic frequency (cm$^{-1}$). The major configurations have a common part ($14\sigma_{1/2}^2 15\sigma_{1/2}^2 9\pi_{1/2}^2 9\pi_{3/2}^2$).}\label{tbl:groundstate}
\begin{tabular}{p{0.8cm} p{0.8cm}  p{0.8cm}   p{7cm}       p{1cm}    p{0.8 cm}   p{0.8 cm}    } \hline\hline
$\Omega$  &$\langle s_z \rangle$ &$\langle l_z \rangle$ &Major configuration & $R$                & $T$&$\omega$      \\\hline
\multicolumn{7}{c}{SD8(13in9)SD, DHF-CI }\\
1/2 &-0.39&0.89&$4\delta_{3/2}^2 4\delta_{5/2}^2 10\pi_{1/2}^2 16(17)^{a}\sigma_{1/2}^1$ ($^{4}\Sigma^-$) 46\%     & 1.674&0    &858 \\
    &     &    &$4\delta_{3/2}^2 4\delta_{5/2}^2 10\pi_{1/2(3/2)}^1 16\sigma_{1/2}^2$ ($^{2}\Pi$) 15\%       &      &     &    \\
    &     &    &$4\delta_{3/2}^1 4\delta_{5/2}^2 10\pi_{1/2}^1 10\pi_{3/2}^1 16\sigma_{1/2}^2$ ($^{2}\Delta$) 1\% &      &     &    \\ [+1 ex]
5/2 &0.50&2.00 &$4\delta_{3/2}^2 4\delta_{5/2}^1 10\pi_{1/2(3/2)}^2 16\sigma_{1/2}^2$ ($^{2}\Delta$) 42\% &1.712&79&974 \\
    &    &     &$4\delta_{3/2}^2 4\delta_{5/2}^1 10\pi_{1/2(3/2)}^2 16\sigma_{1/2}^1 17\sigma_{1/2}^1$ ($^{2}\Delta$)14\% &    &  &    \\
    &    &     &$4\delta_{3/2}^2 4\delta_{5/2}^1 10\pi_{1/2}^1 10\pi_{3/2}^2 16\sigma_{1/2}^1$ ($^{4}\Pi$) 2\% &    &  &    \\
    &    &     &$4\delta_{3/2}^2 4\delta_{5/2}^1 10\pi_{1/2}^1 10\pi_{3/2}^2 16\sigma_{1/2}^1$ ($^{2}\Phi$) 1\% &    &  &    \\ [+1 ex]
7/2 &1.49&2.01 &$4\delta_{3/2}^2 4\delta_{5/2}^1 10\pi_{1/2}^1 10\pi_{3/2}^1 16\sigma_{1/2}^2$ ($^{4}\Delta$) 38\%&1.712&125  &884 \\
    &    &     &$4\delta_{3/2}^2 4\delta_{5/2}^1 10\pi_{1/2}^1 10\pi_{3/2}^1 16\sigma_{1/2}^1 17\sigma_{1/2}^1$($^4\Delta$)13\%&      &     &    \\
    &    &     &$4\delta_{3/2}^2 4\delta_{5/2}^1 10\pi_{1/2}^2 10\pi_{3/2}^1 16(17)\sigma_{1/2}^1$ ($^{2}\Phi$) 17\% &      &     &    \\ [+1 ex]
3/2 &-0.46&1.96&$4\delta_{3/2}^2 4\delta_{5/2}^1 10\pi_{1/2}^1 10\pi_{3/2}^1 16\sigma_{1/2}^2$ ($^{2}\Delta$) 15\% & 1.702&2629 &904 \\
    &    &     &$4\delta_{3/2}^2 4\delta_{5/2}^1 10\pi_{1/2}^1 10\pi_{3/2}^1 16\sigma_{1/2}^1 17\sigma_{1/2}^1$($^{2}\Delta$)5\% &      &     &    \\
    &    &     &$4\delta_{3/2}^2 4\delta_{5/2}^1 10\pi_{1/2}^2 10\pi_{3/2}^1 16(17)\sigma_{1/2}^1$ ($^{2}\Pi$) 10\% &  & &  \\
    &     &    &$4\delta_{3/2}^2 4\delta_{5/2}^2 10\pi_{1/2}^1 10\pi_{3/2}^1 16(17)\sigma_{1/2}^1$($^{4}\Sigma$)18\% &      &     &    \\
    &     &    &$4\delta_{3/2}^1 4\delta_{5/2}^2 10\pi_{1/2}^2 16\sigma_{1/2}^2$($^{2}\Delta$)8\% &      &     &    \\         \hline
\multicolumn{7}{c}{SD8(13in9)SD, MCSCF-CI }\\
1/2 &-0.55&0.94&$4\delta_{3/2}^2 4\delta_{5/2}^2 10\pi_{1/2}^2 16(17)\sigma_{1/2}^1$ ($^{4}\Sigma^-$) 48\% & 1.670& 0   &983 \\
    &     &    &$4\delta_{3/2}^2 4\delta_{5/2}^2 10\pi_{1/2}^1 16\sigma_{1/2}^2$ ($^{2}\Pi$) 13\% &      &     &    \\
    &     &    &$4\delta_{3/2}^2 4\delta_{5/2}^2 10\pi_{1/2}^1 16\sigma_{1/2}^1 17\sigma_{1/2}^1$ ($^{2}\Pi$)4.7\%&      &     &    \\ [+1 ex]
5/2 &0.50 &2.00&$4\delta_{3/2}^2 4\delta_{5/2}^1 10\pi_{1/2(3/2)}^2 16\sigma_{1/2}^2$ ($^{2}\Delta$) 41\% &1.691& 58&1022\\
    &     &    &$4\delta_{3/2}^2 4\delta_{5/2}^1 10\pi_{1/2(3/2)}^2 16\sigma_{1/2}^1 17\sigma_{1/2}^1$ ($^{2}\Delta$)8\% &    &   &    \\
    &     &    &$4\delta_{3/2}^2 4\delta_{5/2}^1 10\pi_{1/2}^1 10\pi_{3/2}^2 16\sigma_{1/2}^1$ ($^{4}\Pi$) 2\%    &      &     &    \\
    &     &    &$4\delta_{3/2}^2 4\delta_{5/2}^1 10\pi_{1/2}^1 10\pi_{3/2}^2 16\sigma_{1/2}^1$ ($^{2}\Phi$) 1\%   &      &     &       \\ [+1 ex]
7/2 &1.47 &2.03&$4\delta_{3/2}^2 4\delta_{5/2}^1 10\pi_{1/2}^1 10\pi_{3/2}^1 16\sigma_{1/2}^2$ ($^{4}\Delta$) 39\%&1.702& 91  &1001\\
    &     &    &$4\delta_{3/2}^2 4\delta_{5/2}^1 10\pi_{1/2}^1 10\pi_{3/2}^1 16\sigma_{1/2}^1 17\sigma_{1/2}^1$($^4\Delta$)13\% &     &    &   \\
    &     &    &$4\delta_{3/2}^2 4\delta_{5/2}^1 10\pi_{1/2}^2 10\pi_{3/2}^1 16(17)\sigma_{1/2}^1$ ($^{2}\Phi$) 18\%  &     &    &   \\ [+1 ex]
3/2 &1.39 &0.11&$4\delta_{3/2}^2 4\delta_{5/2}^1 10\pi_{1/2}^1 10\pi_{3/2}^1 16\sigma_{1/2}^2$ ($^{2}\Delta$) 18\% & 1.681&2632&954 \\
    &     &    &$4\delta_{3/2}^2 4\delta_{5/2}^1 10\pi_{1/2}^1 10\pi_{3/2}^1 16\sigma_{1/2}^1 17\sigma_{1/2}^1$($^{2}\Delta$)7\% &      &     &    \\
    &     &    &$4\delta_{3/2}^2 4\delta_{5/2}^1 10\pi_{1/2}^2 10\pi_{3/2}^1 16(17)\sigma_{1/2}^1$ ($^{2}\Pi$) 12\%  &      &     &    \\
    &     &    &$4\delta_{3/2}^2 4\delta_{5/2}^2 10\pi_{1/2}^1 10\pi_{3/2}^1 16(17)\sigma_{1/2}^1$($^{4}\Sigma$)11\% &      &     &    \\
    &     &    &$4\delta_{3/2}^1 4\delta_{5/2}^2 10\pi_{1/2}^2 16\sigma_{1/2}^2$($^{2}\Delta$)8\% &      &     &    \\      \hline
\multicolumn{7}{c}{SD20(15in10)SD, MCSCF-CI }\\
1/2 &-0.49&0.99&$4\delta_{3/2}^2 4\delta_{5/2}^2 10\pi_{1/2}^2 16(17)\sigma_{1/2}^1$ ($^{4}\Sigma^-$) 52\%& 1.681&0  &1028 \\
    &     &    &$4\delta_{3/2}^2 4\delta_{5/2}^2 10\pi_{1/2(3/2)}^1 16\sigma_{1/2}^2$ ($^{2}\Pi$) 12\%  &      &   &    \\[+1 ex]
5/2 &0.55&1.95&$4\delta_{3/2}^2 4\delta_{5/2}^1 10\pi_{1/2}^2 16\sigma_{1/2}^2$ ($^{2}\Delta$)41\%&1.712 &234 &903   \\
    &    &    &$4\delta_{3/2}^2 4\delta_{5/2}^1 10\pi_{1/2(3/2)}^2 16\sigma_{1/2}^1 17\sigma_{1/2}^1$ ($^{2}\Delta$)17\% &    &  &    \\
    &    &    &$4\delta_{3/2}^2 4\delta_{5/2}^1 10\pi_{1/2}^1 10\pi_{3/2}^2 16\sigma_{1/2}^1$ ($^{4}\Pi$) 2\% &    &  &  \\
    &    &    &$4\delta_{3/2}^2 4\delta_{5/2}^1 10\pi_{1/2}^1 10\pi_{3/2}^2 16\sigma_{1/2}^1$ ($^{2}\Phi$) 1\% &    &  &  \\[+1 ex]
7/2 &1.46 &2.04&$4\delta_{3/2}^2 4\delta_{5/2}^1 10\pi_{1/2}^1 10\pi_{3/2}^1 16\sigma_{1/2}^2$ ($^{4}\Delta$) 43\%&1.712 &260 &954   \\
    &     &    &$4\delta_{3/2}^2 4\delta_{5/2}^1 10\pi_{1/2}^1 10\pi_{3/2}^1 16\sigma_{1/2}^1 17\sigma_{1/2}^1$($^4\Delta$)14\%&      &  &  \\
    &     &    &$4\delta_{3/2}^2 4\delta_{5/2}^1 10\pi_{1/2}^2 10\pi_{3/2}^1 16(17)\sigma_{1/2}^1$ ($^{2}\Phi$) 15\% &      &   &   \\    [+1 ex]
3/2 &-0.38&1.88&$4\delta_{3/2}^2 4\delta_{5/2}^1 10\pi_{1/2}^1 10\pi_{3/2}^1 16\sigma_{1/2}^2$ ($^{2}\Delta$) 13\%   &1.702 &3300 &934\\
    &    &    &$4\delta_{3/2}^2 4\delta_{5/2}^1 10\pi_{1/2}^1 10\pi_{3/2}^1 16\sigma_{1/2}^1 17\sigma_{1/2}^1$($^{2}\Delta$)5\% &  &  & \\
    &    &    &$4\delta_{3/2}^2 4\delta_{5/2}^1 10\pi_{1/2}^2 10\pi_{3/2}^1 16(17)\sigma_{1/2}^1$ ($^{2}\Pi$) 10\%   &   &   &   \\
    &    &    &$4\delta_{3/2}^2 4\delta_{5/2}^2 10\pi_{1/2}^1 10\pi_{3/2}^1 16(17)\sigma_{1/2}^1$($^{4}\Sigma$)22\% &      &  &     \\
    &    &    &$4\delta_{3/2}^1 4\delta_{5/2}^2 10\pi_{1/2}^2 16\sigma_{1/2}^2$($^{2}\Delta$)8\% &      &      \\      \hline\hline
\multicolumn{6}{l}{$^{a}$ The spinor in the bracelet also exists but with less than 5\% composition. }
\end{tabular}
\end{table*}
\end{spacing}

As mentioned above, the previously theoretical predictions of the ground state disagree with the experimental results. The DFT \cite{Citra-JPCA-1999} and RASSI \cite{Suo-JPB-2012} calculations predicted the $^4\Sigma^-$ and $^4\Delta_{7/2}$ ground states, respectively, whereas the experiment results present the $^2\Delta_{5/2}$ ground state. Our calculation gives the ground state of $\Omega$=1/2, which is inconsistent with the RASSI results and also does not support experimental assignment. However, the first excited state is predicted as $\Omega$=5/2, which only locates 234 cm$^{-1}$ above the ground state on our highest level of calculation. In addition, our predicted energy separation between $\Omega$=5/2 and 7/2 is only several tens cm$^{-1}$, which is far less than the RASSI value of 1327 cm$^{-1}$ \cite{Suo-supplement}. The 1/2, 5/2 and 7/2 are so close that the ground state can be easily polluted by two other low-lying excited states. As can be seen in Table \ref{tbl:groundstate}, the equilibrium bond length of 5/2 state is 1.712 \AA \ that is only 0.012 \AA \ shorter than Pang \textit{et al.}'s result \cite{Pang-JPCA-2010}. The harmonic vibrational frequency 903 $cm^{-1}$ agrees well with the experimental value of 909 $cm^{-1}$ obtained by Pang and co-workers \cite{Pang-JPCA-2010}. Therefore, the 5/2 state in our calculation should be the low-lying state that contributes to the experimental spectra observed by Pang \textit{et al.} and Adam \textit{et al.} \cite{Adam-JMS-2013}.

\begin{spacing}{1}
\begin{table*}\small
\caption{The lowest-lying electronic states of CoO and RhO for $\Omega$=1/2, 3/2, 5/2, and 7/2 symmetries, as calculated at the KRCI SD8(13in9)SD level in the MCSCF+CI implementations, where $\langle s_z \rangle$ and $\langle l_z \rangle$ are expectation values of $s_z$ and $l_z$ operators, and $R$, $T$ are the equilibrium bond length ({\AA}) and excited energy (cm$^{-1}$). The major configuration of CoO and RhO have the common parts ($7\sigma_{1/2}^2 8\sigma_{1/2}^2 3\pi_{1/2}^2 3\pi_{3/2}^2$) and ($10\sigma_{1/2}^2 11\sigma_{1/2}^2 5\pi_{1/2}^2 5\pi_{3/2}^2$), respectively.}\label{tbl:CoORhO}
\begin{tabular}{p{0.8cm} p{0.8cm}  p{0.8cm}   p{7cm}       p{1cm}     p{0.8 cm}     }\hline\hline
              $\Omega$  &$\langle s_z \rangle$&$\langle l_z \rangle$ &Major configuration & $R$ & $T$    \\ \hline
\multicolumn{6}{c}{ CoO }\\
7/2 &1.50&2.00&$1\delta_{3/2}^2 1\delta_{5/2}^1 4\pi_{1/2}^1 4\pi_{3/2}^1 9(10)^{a}\sigma_{1/2}^2$ ($^{4}\Delta$) 65\%&1.620 &0    \\   [+2 ex]
5/2 &0.50&2.00&$1\delta_{3/2}^2 1\delta_{5/2}^1 4\pi_{1/2(3/2)}^2 9\sigma_{1/2}^2$ ($^{2}\Delta$)31\%&1.620 &290   \\
    &    &    &$1\delta_{3/2}^2 1\delta_{5/2}^1 4\pi_{1/2(3/2)}^2 9\sigma_{1/2}^1 10\sigma_{1/2}^1$ ($^{2}\Delta$)23\% &    &    \\[+2 ex]
3/2 &-0.50&2.00&$1\delta_{3/2}^2 1\delta_{5/2}^1 4\pi_{1/2}^1 4\pi_{3/2}^1 9\sigma_{1/2}^2$ ($^{2}\Delta$) 16\%  &1.620 &670  \\
    &    &    &$1\delta_{3/2}^2 1\delta_{5/2}^1 4\pi_{1/2}^1 4\pi_{3/2}^1 9\sigma_{1/2}^1$($^{2}\Delta)$13\% &   &     \\[+2 ex]
1/2 &-1.50&1.98&$1\delta_{3/2}^1 1\delta_{5/2}^2 4\pi_{1/2}^1 4\pi_{3/2}^1 9\sigma_{1/2}^2$ ($^{4}\Delta$) 42\%& 1.610&786   \\
    &    &     &$1\delta_{3/2}^1 1\delta_{5/2}^2 4\pi_{1/2}^1 4\pi_{3/2}^1 9\sigma_{1/2}^1 10\sigma_{1/2}^1$ ($^{4}\Delta$) 35\%&  &   \\\hline
\multicolumn{6}{c}{RhO}\\
3/2 &1.49&0.008&$2\delta_{3/2}^2 2\delta_{5/2}^2 6\pi_{1/2}^1 6\pi_{3/2}^1 12(13)\sigma_{1/2}^1$ ($^{4}\Sigma^-$) 76\%&1.712 &0 \\
    &    &     &$2\delta_{3/2}^2 2\delta_{5/2}^2 6\pi_{1/2}^2 6\pi_{3/2}^1 $ ($^{2}\Pi$) 2\%                       &      &       \\ [+2 ex]
1/2 &0.49&0.01 &$2\delta_{3/2}^2 2\delta_{5/2}^2 6\pi_{1/2}^2 12(13)\sigma_{1/2}^1$ ($^{2}\Sigma^-$) 32\%&1.712 &223   \\
    &    &     &$2\delta_{3/2}^2 2\delta_{5/2}^2 6\pi_{3/2}^2 12(13)\sigma_{1/2}^1$ ($^{2}\Sigma^-$) 21\%&      &         \\ [+2 ex]
7/2 &1.49&2.01&$2\delta_{3/2}^2 2\delta_{5/2}^1 6\pi_{1/2}^1 6\pi_{3/2}^1 12\sigma_{1/2}^2$ ($^{4}\Delta$) 66\%     &1.778 &3793   \\
    &    &    &$2\delta_{3/2}^2 2\delta_{5/2}^1 6\pi_{1/2}^1 6\pi_{3/2}^1 12\sigma_{1/2}^1 13\sigma_{1/2}^1$($^4\Delta$)5\%&  &    \\
    &    &    &$2\delta_{3/2}^2 2\delta_{5/2}^1 6\pi_{1/2}^2 6\pi_{3/2}^1 12\sigma_{1/2}^1$ ($^{2}\Phi$) 2\% &      &     \\[+2 ex]
5/2 &0.50&2.00&$2\delta_{3/2}^2 2\delta_{5/2}^1 6\pi_{1/2(3/2)}^2 12\sigma_{1/2}^2$ ($^{2}\Delta$)35\% &1.740 &3976 \\
    &    &    &$2\delta_{3/2}^2 2\delta_{5/2}^1 6\pi_{1/2(3/2)}^2 12\sigma_{1/2}^1 13\sigma_{1/2}^1$ ($^{2}\Delta$)15\% &    &     \\
    &    &    &$2\delta_{3/2}^2 2\delta_{5/2}^1 6\pi_{1/2}^1 6\pi_{3/2}^2 12\sigma_{1/2}^1$ ($^{4}\Pi$) 2\% &    &     \\
    &    &    &$2\delta_{3/2}^2 2\delta_{5/2}^1 6\pi_{1/2}^1 6\pi_{3/2}^2 12\sigma_{1/2}^1$ ($^{2}\Phi$) 1\% &    &     \\ [+2 ex]\hline\hline
\multicolumn{6}{l}{$^{a}$ The spinor in the bracelet also exists but with less than 5\% composition. }
\end{tabular}
\end{table*}
\end{spacing}


The strongly relativistic effect leads the electronic states of IrO could only be identified by their $\Omega$ values, not by $\lambda$ value. However, the expectation values $\langle s_z\rangle=\Sigma_{i}m_s$ and $\langle l_z\rangle=\Sigma_{i}m_l$ suggest the major component in the $\Lambda-S$ notation. Here, we also provide the corresponding $\Lambda-S$ notation for each configuration in Table \ref{tbl:groundstate}. As illustrated by Table \ref{tbl:groundstate}, the lowest lying states of $\Omega$=1/2, 3/2, 5/2, and 7/2 have a common electronic occupation of $14\sigma_{1/2}^2 15\sigma_{1/2}^2 9\pi_{3/2}^2 9\pi_{1/2}^2$ in valance orbitals. The $14\sigma_{1/2}$ is mainly $2s$ atomic orbital of O. The $15\sigma_{1/2}$ is bonding orbital of Ir$(5d_{\sigma})$ + O$(2p_{\sigma})$. The $9\pi_{1/2}$ and $9\pi_{3/2}$ are two components of bonding $\pi$ orbital (Ir$(5d_{\pi})$+O$(2p_{\pi})$). These orbitals are low in energy and are fully occupied in all states in present study. Remaining six spinors in valence space, namely $16\sigma_{1/2}$, $17\sigma_{1/2}$, $4\delta_{5/2}$, $4\delta_{3/2}$ and $10\pi_{1/2}$ and $10\pi_{3/2}$ are occupied by 7 electrons in flexible ways that could generate many electronic states with relatively similar energies. Two components of the $\delta$ spinor are mainly composed of Ir $5d_{\delta}$ atomic orbital that is non-bonding. Two $10\pi$ spinors are anti-bonding combination of Ir $5d_{\pi}$ and O $2p_{\pi}$. The $16\sigma$ is mainly composed of Ir $6s^2$  and $17\sigma$ is anti-bonding of Ir $5d_{\sigma}$ and O $2p_{\sigma}$.

The ground state $\Omega$=1/2 has the leading configuration of $4\delta_{3/2}^2 4\delta_{5/2}^2 10\pi_{1/2}^2 16\sigma_{1/2}^1$ (or $17\sigma_{1/2}^1$ with composition less than 5\%), which corresponds to $^{4}\Sigma^-$ with $\delta^4\pi^2\sigma^1$ molecular orbital occupation. The configuration $4\delta_{3/2}^2 4\delta_{5/2}^2 10\pi_{1/2}^1 16\sigma_{1/2}^2$ corresponding to $^{2}\Pi$ ($\delta^4\pi^1\sigma^2$) also contributes to the lowest $\Omega$=1/2 state but with smaller component. Comparing with predictions of RASSI calculation, it suggests that the 1/2 state is mixture of $^2\Pi, ^4\Delta, ^4\Sigma^-$ components. The lowest lying state of $\Omega$=5/2 has the leading configuration $4\delta_{3/2}^2 4\delta_{5/2}^1 10\pi_{1/2}^2 16\sigma_{1/2}^2$, corresponding to $^2\Delta$. Therefore, the present result supports the previous assignment that the $\Omega$=5/2 state is arising from the $^2\Delta$ state in Pang's work \cite{Pang-JPCA-2010}. Nevertheless, our results show that the lowest $\Omega$=5/2 state has $\delta^3\pi^2\sigma^2$ molecular orbital occupation, not $\delta^3\pi^4$ as supposed by Pang and co-workers \cite{Pang-JPCA-2010}. The lowest lying state of $\Omega$=7/2 has a major configuration $4\delta_{3/2}^2 4\delta_{5/2}^1 10\pi_{1/2}^1 10\pi_{3/2}^1 16\sigma_{1/2}^2$, which corresponds to $^{4}\Delta$ if described in the $\Lambda-S$ notation. Therefore, the lowest lying state of $\Omega$=7/2 comes mainly from the $\delta^3\pi^2\sigma^2$ occupation, which is consistent with the RASSI results for the $^4\Delta_{7/2}$. By analyzing the energy separations and leading configurations for the lowest $\Omega$=1/2, 5/2, and 7/2 states, we suggest that the none-bonding $16\sigma$, $4\delta$ and anti-bonding $10\pi$ orbitals are close in energy and $17\sigma$ is slightly higher. This is consistent with previous CASPT2 result that has included scaler relativistic effect in calculation.

It is interesting to compare IrO with its iso-valent molecules CoO and RhO. It already known that CoO and RhO have the $^4\Delta$ and $^4\Sigma^-$ ground states, respectively. Here, we have calculated these two molecules at KRCI level and the results of the low-lying states are summarized in Table. \ref{tbl:CoORhO}. For CoO, the $\Omega$=7/2 is assigned as the ground state, which has the leading configuration $1\delta_{3/2}^2 1\delta_{5/2}^1 4\pi_{1/2}^1 4\pi_{3/2}^1 9(10)\sigma_{1/2}^2$ ($^{4}\Delta$). The first excited state is $\Omega$=5/2, which corresponds to a $^2\Delta$ state. The adiabatic excitation energy is 290 cm$^{-1}$, being closed to the previous experimental value of $244$ cm$^{-1}$ \cite{CoO1987}. For RhO, our calculations show that the lowest-lying state has the $\Omega$ value of 3/2, which is mainly contributed by the leading configuration $2\delta_{3/2}^2 2\delta_{5/2}^2 6\pi_{1/2}^1 6\pi_{3/2}^1 12(13)\sigma_{1/2}^1$ ($^{4}\Sigma^-$). Our results of the ground state of CoO and RhO are in agreement with the previous experimental and theoretical studies \cite{CoO1979,CoO1987,CoO1993,CoO1997,CoO2005,CoO2006, RhO1998, RhO1999, RhO2002, RhO2003,RhO2005,RhO2007,RhO2009}. The low-lying states of three molecules present different characters. For instance, both of the CoO and IrO molecules have very dense states in ground state region. The energy gaps of four lowest states of CoO are only several hundreds cm$^{-1}$ and are much closer than corresponding values in RhO. For RhO, four lowest states are divided into two groups. As illustrated in Table \ref{tbl:CoORhO}, the 3/2 and 1/2 states are closed in energy and the 7/2 and 5/2 are nearly degenerate. However, the 7/2 and 5/2 states are well separated with the 1/2 and 3/2 states and locate more than 3500 $cm^{-1}$ higher. Thus, it is easier to assign the ground state of RhO than CoO. For the IrO molecule, it is much trouble because all of the three lowing states 1/2, 5/2 and 7/2 locate within 250 cm$^{-1}$, as shown in our calculation. Such near-degeneracy causes great difficulty in identification of the ground state of IrO.

Tracing back to the ground states of the group-9 elements Co, Rh and Ir, both the ground states of Co and Ir have a $d^7s^2$ configuration, whereas the Rh atom prefers a $d^8s^1$ configuration, which suggests the smaller electronic repulsion of d shells of the Rh atom. Provided that the group-9 metals bonded with oxygen atom somehow do not change the occupation numbers of those none-bonding $\sigma$ and $\delta$ orbitals, one can expect that IrO should be similar to CoO instead of RhO. However, our results show that the strongly relativistic effect of the Ir atom should be taken into account during such comparison. First of all, the relativistic effect causes the contraction of the $s$ and $p$ orbitals and then the increasing screening effect, which yields the expansion of the $d$ orbital. Thus, the repulsion of d electrons in the Ir atom should smaller than that in the Co atom. Also, the non-bonding $\delta$, $\sigma$ and anti-bonding $\pi$ orbitals are much closer in IrO than CoO, which leads to denser low-lying states of IrO than CoO. When the multiply-degeneracy of $d$ open shell of Ir element prevails over the $d$-shell electron repulsion effect, it will give rise to a low spin state, i.e., $\Omega$=1/2, of the IrO molecule in which $\delta$ orbital is fully occupied. In addition, these non-bonding orbitals are so close that it is easily to put electron in none-bonding $\sigma$(Ir 6s) orbital due to the electron repulsion in $\delta$ (Ir $5d_{\delta}$) orbital \cite{Eugen-JCE-2010}, which results in the high spin state, $\Omega=$7/2 and 5/2, of IrO close to the ground-state just like their counter part component of CoO.

\subsection{Excited states and assignment of the experimental spectra}

\begin{spacing}{1}
\begin{table*}\small
\caption{The low-lying states of IrO with $\Omega=5/2$, as calculated at the KRCI SD8(13in9)SD level with spinor obtained by KR-MCSCF implementations, where $\langle s_z \rangle$ and $\langle l_z \rangle$ are expectation values of $s_z$ and $l_z$ operators, $R$ is equilibrium bond length (\AA), $T^{Adi}_{cv2z}$ and $T^{Com}$ are excited energy (cm$^{-1}$), and $f$ is oscillation strength (a.u.). The major configurations have a common part ($14\sigma_{1/2}^2 15\sigma_{1/2}^2 9\pi_{1/2}^2 9\pi_{3/2}^2$).}\label{grid_mlmmh}
\begin{tabular}{p{0.9cm} p{0.9cm}  p{0.9cm}  p{0.9cm}  p{6cm}       p{1cm}     p{1cm}    p{1.1cm}  } \hline\hline
State&$\langle s_z \rangle$ &$\langle l_z \rangle$&$R$&Major configuration &$T^{Adi}_{cv2z}$ & $T^{Com}$ & $f$\\\hline
0 &0.50 &2.00&1.712&$4\delta_{3/2}^2 4\delta_{5/2}^1 10\pi_{1/2}^2 16\sigma_{1/2}^2$  41\%                 &0    & 0   &   \\
  &     &    &     &$4\delta_{3/2}^2 4\delta_{5/2}^1 10\pi_{1/2}^2 16\sigma_{1/2}^1 17\sigma_{1/2}^1$  15\%&     &     &   \\[+2 ex]
1 &0.50 &2.00&1.712&$4\delta_{3/2}^2 4\delta_{5/2}^2 10\pi_{1/2}^1 10\pi_{3/2}^1 16\sigma_{1/2}^1$ 19\%    &8481 &8373 &0.0003  \\
  &     &    &     &$4\delta_{3/2}^2 4\delta_{5/2}^2 10\pi_{1/2}^1 10\pi_{3/2}^1 17\sigma_{1/2}^1$  8\%    &     &     &   \\[+1 ex]
2 &0.52 &1.98&1.774&$4\delta_{3/2}^2 4\delta_{5/2}^1 10\pi_{1/2}^2 16\sigma_{1/2}^1 17\sigma_{1/2}^1$ 19\% &10350&11612&0.0003\\
  &	    &	 &     &$4\delta_{3/2}^2 4\delta_{5/2}^1 10\pi_{1/2}^1 10\pi_{3/2}^2 16\sigma_{1/2}^1 $ 12\%   &	 &     &      \\[+1 ex]
3 &1.47 &1.03&1.787&$4\delta_{3/2}^2 4\delta_{5/2}^1 10\pi_{1/2}^1 10\pi_{3/2}^2 16\sigma_{1/2}^1$  12\%   &12691&13462&0.0003\\
  &     &    &     &$4\delta_{3/2}^2 4\delta_{5/2}^1 10\pi_{1/2}^1 16\sigma_{1/2}^2 17\sigma_{1/2}^1$  7\% &     &     &       \\[+2 ex]
4 &1.72 &0.78&1.774&$4\delta_{3/2}^2 4\delta_{5/2}^1 10\pi_{1/2}^1 16\sigma_{1/2}^2 17\sigma_{1/2}^1$ 18\% &13893&14385&0.0021 \\
  &	    &	 &     &$4\delta_{3/2}^2 4\delta_{5/2}^1 10\pi_{1/2}^1 10\pi_{3/2}^2 16\sigma_{1/2}^1$ 12\%    &     &     &       \\[+2 ex]
5 &0.46 &2.04&1.787&$4\delta_{3/2}^2 4\delta_{5/2}^2 10\pi_{1/2}^1 10\pi_{3/2}^1 16\sigma_{1/2}^1$ 12\%    &14723&15495&0.0001\\
  &	    &	 &     &$4\delta_{3/2}^2 4\delta_{5/2}^2 10\pi_{1/2}^1 10\pi_{3/2}^1 17\sigma_{1/2}^1$ 10\%    &     &     &       \\[+2 ex]
6 &-0.46&2.96&1.787&$4\delta_{3/2}^2 4\delta_{5/2}^1 10\pi_{1/2}^1 16\sigma_{1/2}^2 17\sigma_{1/2}^1$ 10\% &15448&16483&0.0054\\
  &	    &	 &     &$4\delta_{3/2}^1 4\delta_{5/2}^2 10\pi_{1/2}^1 10\pi_{3/2}^2 16\sigma_{1/2}^1$ 11\%    &     &     &       \\
  &     &    &     &$4\delta_{3/2}^2 4\delta_{5/2}^1 10\pi_{1/2}^1 10\pi_{3/2}^2 16\sigma_{1/2}^1$ 7\%     &     &     &	   \\[+2 ex]
7 &-0.48&2.98&1.787&$4\delta_{3/2}^2 4\delta_{5/2}^1 10\pi_{1/2}^1 16\sigma_{1/2}^2 17\sigma_{1/2}^1$ 22\% &17231&17694&0.0019\\
  &	    &	 &     &$4\delta_{3/2}^2 4\delta_{5/2}^1 10\pi_{3/2}^2 16\sigma_{1/2}^2$      6\%              &     &     &       \\
  &     &    &     &$4\delta_{3/2}^2 4\delta_{5/2}^1 10\pi_{1/2}^2 10\pi_{3/2}^1 16\sigma_{1/2}^1$    5\%  &     &     &	   \\[+2 ex]
8 &-0.48&2.98&1.787&$4\delta_{3/2}^2 4\delta_{5/2}^1 10\pi_{1/2}^1 16\sigma_{1/2}^2 17\sigma_{1/2}^1$ 13\% &18316&19304&0.0145\\
  &     &	 &     &$4\delta_{3/2}^1 4\delta_{5/2}^1 10\pi_{1/2}^2 10\pi_{3/2}^1 16\sigma_{1/2}^2$    11\% &     &     &       \\[+2 ex]
9 &0.48 &2.02&1.787&$4\delta_{3/2}^1 4\delta_{5/2}^1 10\pi_{1/2}^2 10\pi_{3/2}^1 16\sigma_{1/2}^2$ 16\%    &20043&21035&0.0047\\
  &     &    &     &$4\delta_{3/2}^1 4\delta_{5/2}^1 10\pi_{1/2}^2 10\pi_{3/2}^1 16\sigma_{1/2}^1 17\sigma_{1/2}^1$ 5\% &    &    &    \\
  &     &    &     &$4\delta_{3/2}^2 4\delta_{5/2}^1 10\pi_{3/2}^2 16\sigma_{1/2}^2$ 5\%                    &    &     &        \\
  &	    &	 &     &$4\delta_{3/2}^2 4\delta_{5/2}^1 10\pi_{1/2}^1 10\pi_{3/2}^2 16\sigma_{1/2}^1$ 5\%      &    &     &		\\
10&0.52 &1.98&1.794&$4\delta_{3/2}^2 4\delta_{5/2}^1 10\pi_{3/2}^2 16\sigma_{1/2}^2$ 7\%                   &22022&22575&0.0217\\
  &	    &	 &     &$4\delta_{3/2}^2 4\delta_{5/2}^1 10\pi_{3/2}^1 16\sigma_{1/2}^2 17\sigma_{1/2}^1$ 6\%  &     &     &     \\
  &	    &	 &     &$4\delta_{3/2}^2 4\delta_{5/2}^1 10\pi_{1/2}^2 16\sigma_{1/2}^1 17\sigma_{1/2}^1$ 3\%  &     &     &     \\
  &	    &	 &     &$4\delta_{3/2}^2 4\delta_{5/2}^2 10\pi_{3/2}^1 16\sigma_{1/2}^1 17\sigma_{1/2}^1$ 7\%  &     &     &     \\  [+2 ex]
11&-0.50&3.00&1.787&$4\delta_{3/2}^1 4\delta_{5/2}^2 10\pi_{1/2}^1 10\pi_{3/2}^2 16\sigma_{1/2}^1$ 10\%    &22320&22933&0.0069\\
  &     &    &     &$4\delta_{3/2}^2 4\delta_{5/2}^1 10\pi_{1/2}^1 10\pi_{3/2}^1 16\sigma_{1/2}^1 17\sigma_{1/2}^1$ 8\% &    &    &   \\[+2 ex]
12&1.41 &1.09&1.787&$4\delta_{3/2}^2 4\delta_{5/2}^2 10\pi_{3/2}^1 16\sigma_{1/2}^1 17\sigma_{1/2}^1$ 29\% &22913&24426&$<10^{-4}$ \\[+2 ex]
13&0.52 &1.98&1.787&$4\delta_{3/2}^1 4\delta_{5/2}^2 10\pi_{1/2}^2 16\sigma_{1/2}^1 17\sigma_{1/2}^1$ 12\% &23518&24708&0.034\\
  &     &    &     &$4\delta_{3/2}^2 4\delta_{5/2}^2 10\pi_{1/2}^1 16\sigma_{1/2}^1 17\sigma_{1/2}^1$  6\% &     &     &   \\
  &     &    &     &$4\delta_{3/2}^2 4\delta_{5/2}^1 10\pi_{1/2}^1 10\pi_{3/2}^1 16\sigma_{1/2}^1 17\sigma_{1/2}^1$11\% &  &    &   \\[+2 ex]
14&2.38 &0.12&1.808&$4\delta_{3/2}^1 4\delta_{5/2}^1 10\pi_{1/2}^1 16\sigma_{1/2}^2 17\sigma_{1/2}^1$ 30\% &24796&26037&0.0005\\[+2 ex]
15&-0.21&2.71&1.808&$4\delta_{3/2}^1 4\delta_{5/2}^2 10\pi_{1/2}^2 10\pi_{3/2}^1 16\sigma_{1/2}^2$ 9\%     &25324&26761&0.0106\\[+2 ex] \hline\hline
\end{tabular}
\end{table*}
\end{spacing}

\begin{spacing}{1}
\begin{table*}\small
\caption{The low-lying states of IrO with $\Omega=7/2$, as calculated at the KRCI SD8(13in9)SD level with spinors obtained by KR-MCSCF, where $\langle s_z \rangle$ and $\langle l_z \rangle$ are expectation values of $s_z$ and $l_z$ operators, $R$ is equilibrium bond length (\AA), $T^{Adi}_{cv2z}$ and $T^{Com}$ are excited energy (cm$^{-1}$), and $f$ is oscillation strength (a.u.). The major configurations have a common part ($14\sigma_{1/2}^2 15\sigma_{1/2}^2 9\pi_{1/2}^2 9\pi_{3/2}^2$).}\label{grid_mlmmh}
\begin{tabular}{p{0.8cm} p{0.8cm}  p{0.8cm}  p{0.8cm}  p{6cm}       p{1cm}     p{1cm}    p{1.1cm}  } \\ \hline\hline
State&$\langle s_z \rangle$ &$\langle l_z \rangle$&$R$({\AA})&Major configuration &$T^{Adi}_{cv2z}$ & $T^{Com}$   & $f$  \\
\hline
0&1.48 &2.02&1.723&$4\delta_{3/2}^2 4\delta_{5/2}^1 10\pi_{1/2}^1 10\pi_{3/2}^1 16\sigma_{1/2}^2$ 36\%     & 0   &0     &     \\
 &     &    &	  &$4\delta_{3/2}^2 4\delta_{5/2}^1 10\pi_{1/2}^2 10\pi_{3/2}^1 16\sigma_{1/2}^1$ 15\%     &	 &      &     \\[+2 ex]
1&1.49 &2.01&1.776&$4\delta_{3/2}^2 4\delta_{5/2}^1 10\pi_{1/2}^2 16\sigma_{1/2}^1 17\sigma_{1/2}^1$ 11\%  &8662 &9560  &$<10^{-4}$\\
 &     &    &	  &$4\delta_{3/2}^2 4\delta_{5/2}^1 10\pi_{1/2}^1 10\pi_{3/2}^2 16\sigma_{1/2}^1$ 7\%      &	 &      &     \\[+2 ex]
2&0.49 &3.01&1.776&$4\delta_{3/2}^2 4\delta_{5/2}^1 10\pi_{1/2}^2 16\sigma_{1/2}^1 17\sigma_{1/2}^1$ 18\%  &12060&13126 &0.005\\
 &	   &    &     &$4\delta_{3/2}^2 4\delta_{5/2}^1 10\pi_{1/2}^1 10\pi_{3/2}^2 16\sigma_{1/2}^1$ 5\%      &     &      &     \\
 &	   &    &     &$4\delta_{3/2}^2 4\delta_{5/2}^1 10\pi_{3/2}^2 16\sigma_{1/2}^1 17\sigma_{1/2}^1$ 4\%   &	 &      &     \\[+2 ex]
3&-0.43&3.93&1.723&$4\delta_{3/2}^1 4\delta_{5/2}^2 10\pi_{1/2}^1 10\pi_{3/2}^1 16\sigma_{1/2}^2$ 35\%     &15684&15428 &0.0001\\
 &	   &	&     &$4\delta_{3/2}^1 4\delta_{5/2}^2 10\pi_{1/2}^2 10\pi_{3/2}^1 16\sigma_{1/2}^1$ 9\%      &	 &      &    \\[+2 ex]
4&-0.46&3.96&1.776&$4\delta_{3/2}^1 4\delta_{5/2}^2 10\pi_{1/2}^1 10\pi_{3/2}^1 16\sigma_{1/2}^2$ 32\%     &16208&16547 &$<10^{-4}$\\[+2 ex]
5&0.40 &3.10&1.776&$4\delta_{3/2}^2 4\delta_{5/2}^1 10\pi_{1/2}^1 16\sigma_{1/2}^2 17\sigma_{1/2}^1$ 21\%  &17661&18585 &0.0019 \\
 &     &    &	 &$4\delta_{3/2}^2 4\delta_{5/2}^1 10\pi_{1/2}^2 16\sigma_{1/2}^1 17\sigma_{1/2}^1$ 11\%   &	 &      &       \\[+2 ex]
6&1.47&2.03&1.820&$4\delta_{3/2}^2 4\delta_{5/2}^1 10\pi_{1/2}^1 10\pi_{3/2}^1 16\sigma_{1/2}^1 17\sigma_{1/2}^1$ 14\%&20027&21013 &0.0046\\
 &    &    &     &$4\delta_{3/2}^2 4\delta_{5/2}^1 10\pi_{1/2}^2 16\sigma_{1/2}^1 17\sigma_{1/2}^1$ 14\%   &     &      &       \\
 &    &    &     &$4\delta_{3/2}^2 4\delta_{5/2}^1 10\pi_{1/2}^1 16\sigma_{1/2}^2 17\sigma_{1/2}^1$ 8\%    &     &      &       \\[+2 ex]
7&1.48&2.02&1.820&$4\delta_{3/2}^2 4\delta_{5/2}^1 10\pi_{1/2}^1 10\pi_{3/2}^1 16\sigma_{1/2}^1 17\sigma_{1/2}^1$ 11\%&22319&23478 &0.099\\
 &	  &	   &     &$4\delta_{3/2}^2 4\delta_{5/2}^1 10\pi_{1/2}^1 10\pi_{3/2}^1 16\sigma_{1/2}^2$ 11\%      &	 &      &       \\[+2 ex]
8&0.52&2.98&1.820&$4\delta_{3/2}^2 4\delta_{5/2}^1 10\pi_{1/2}^1 16\sigma_{1/2}^2 17\sigma_{1/2}^1$ 16\%   &24497&25546 &0.0035 \\[+2 ex]
9&0.58&2.92&1.820&$4\delta_{3/2}^1 4\delta_{5/2}^2 10\pi_{1/2}^1 10\pi_{3/2}^1 16\sigma_{1/2}^2$ 10\%      &26374&26671 &0.0162 \\
 &    &    &     &$4\delta_{3/2}^1 4\delta_{5/2}^2 10\pi_{1/2}^2 10\pi_{3/2}^1 16\sigma_{1/2}^1$ 12\%      &     &      &	     \\ \hline\hline
\end{tabular}
\end{table*}
\end{spacing}

The calculated excited states for $\Omega$=5/2 and 7/2 are summarized in Table 3 and 4 in order to explain the observed visible transition bands. The adiabatic excited states are calculated at the [dyall.cv2z (Ir) and aug-cc-pCVDZ (O)] basis sets, which yields the equilibrium bond length $R$ and the adiabatic excited energy $T^{Adi}_{cv2z}$. Besides, we also give a composite data of the excited energy $T^{Com}$, as evaluated for larger basis set (See supplemental material for the computational details and the complete data tabular). The oscillator strength $f$ associated to the transition from the lowest state in each $\Omega$ component to the upper states are given along with the excitation energy, which provides a reference to determinate which excited state can be assigned to the observed transition bands. The excited states of IrO exhibit stronger multiconfigurational features and therefore are more difficult to analyze the molecular orbital picture than the ground state. Moreover, our calculation for excited states is mainly based on the small basis set, limited by the huge computational demanding, which is supposed to bring uncertainty of 100-2000 $cm^{-1}$ due to the basis set incompleteness. Therefore, our analysis for the experimental transition band is conducted in a qualitative way.

As shown in Table 3, the 6th to 8th excited states have the excited energies $T$=15448 and 18316 cm$^{-1}$, respectively, that are close to the experimental values [17.6] and [17.8], and also have significant oscillator strengths. As compared with the electronic configurations of the ground state, the 6th to 8th  excited states can be traced back to the electron excitations from $10\pi_{1/2}$ to $17\sigma_{1/2}$. The obtained equilibrium bond length for the 6th to 8th excited states is 1.787 {\AA}, which differs from the experimental value $R$=1.7969 {\AA} by about 0.1 {\AA}. This deviation in the bond length is not surprised for KRCI calculation of excited states. An assessment of accuracy of KRCI is conducted by Stefan, et al., \cite {Stefan-Thesis-2013} which shows that the KRCI calculation performs well for prediction of excited energy but with apparently large deviations in bond length. Therefore, consider the good agreement of the excited energy, the strong oscillation strength, and the clear electronic excitation path, we can suggest that the 6th to 8th excited states could contribute to the observed visible bands [17.6] 2.5-X$^2\Delta_{5/2}$ and [17.8] 2.5-X$^2\Delta_{5/2}$.

The 9th to 11th excited states give the excitation energies of 22043 and 22320 cm$^{-1}$ and the significant oscillator strengths of around 0.0047 - 0.0217. Such excited energies are close to the experimental values of transitions [21.5]-[22.0] 2.5-X2.5. The corresponding bond length is around 1.787 and 1.794 \AA, which is consistent with the experimental value 1.7874 \AA. Be compared with  the electronic configurations of the ground state, the 9th to 11th excited states are mainly arising from electron excitation from $4\delta_{3/2}$ to $10\pi_{3/2}$, $4\delta_{3/2}$ to $4\delta_{5/2}$, and $10\pi_{1/2}$ to $10\pi_{3/2}$. The good agreement with the experimental values suggests that the 9th to 11th excited states can be assigned to the observed transition bands [21.5] 2.5-X$^2\Delta_{5/2}$ and [22.0] 2.5-X$^2\Delta_{5/2}$.

The 13th excited state is arising from electron excitation from $4\delta_{3/2}$ to $4\delta_{5/2}$ and $10\pi_{1/2}$ to $10\pi_{3/2}$, which gives the excited energy $T$=23518 cm$^{-1}$ that is close to the experimental value [23.3]. The corresponding oscillator strength is strong, be around 0.034, which indicates that this state is highly probably observed in experiment. Therefore, we can assign the 13th state to the observed band [23.3] 2.5-2.5.

Compared to the wealth of the experimental spectra of $\Omega$=5/2, only one transition band [21.9] 3.5-3.5 is observed for $\Omega$=7/2 \cite{Pang-JPCA-2010}. As shown in Table. 4, the 6th and 7th excited states give the excited energies around 20027-22318 cm$^{-1}$ that are close to the experimental value [21.9]. Their oscillation lengths are around 0.0046-0.099, indicating that these states are highly probable to be observed in experiments. The corresponding electronic excitations are also clear, mainly arising from singly excitation from $16\sigma_{1/2}$ and $10\pi_{3/2}$ to $17\sigma_{1/2}$. Therefore, we suggest that one of the 6th and 7th excited states with $\Omega$=7/2 may contribute to the experimental visible transition band [21.9] 3.5-3.5.

\section{Conclusions}
In summary, the electronic structure of IrO is calculated by using fully relativistic multi-reference configuration interaction method. Four lowest-lying states for $\Omega$=1/2, 3/2, 5/2, and 7/2 are calculated with the aim to determinate the ground state of IrO. Our results indicate that the ground state of IrO is $\Omega$=1/2, which is highly mixed by component of the $^4\Sigma^-$ and $^2\Pi$ states. Two low-lying states, namely, 5/2 and 7/2, are nearly degenerated with the 1/2 state and locate only 234 and 260 cm$^{-1}$ above. The quite small energy separations among the three $\Omega$ states bring the great difficulty for the identification of the ground state, which causes the discrepancy between the experiments and the theoretical calculations. However, our calculation supports that the low-lying state $5/2$ should contribute to most of the experimental observed spectra. This state has the equilibrium bond length of 1.712 \AA \ and vibrational frequency of 903 cm$^{-1}$, which agrees with the experimental value 1.724 \AA \ and 909 cm$^{-1}$ \cite{Pang-JPCA-2010}. Furthermore, the excited states of IrO are investigated for $\Omega$=5/2 and 7/2 that can be used to interpret the experimental spectra. Six excited states are assigned to the observed six transition bands of LIF spectrum in the visible region, i.e., [17.6] 2.5-$^2\Delta_{5/2}$, [17.8] 2.5-$^2\Delta_{5/2}$, [21.5] 2.5-$^2\Delta_{5/2}$, [22.0] 2.5-$^2\Delta_{5/2}$, [21.9] 3.5-$\Omega$=3.5, and [23.3] 2.5-2.5, through comparing the excited energy values, the oscillator strengths and the possible excitation paths.

Our calculation is implemented at the high level that the relativistic effect and spin-orbit coupling are taken into account at the same foot. The multi-reference properties of 5d shell of Ir is sufficiently considered through KR-MCSCF implementation. The electronic correlation is considered up to the 4f electrons of Ir. The strong multiconfigurational features of the electronic structure of IrO have been clearly demonstrated in our calculations. Our calculation still cannot provide the direct theoretical proof that the ground state is the $\Omega$=5/2 state. This urges more systematic theoretical and experimental work, for example, to give the relative position of the $\Omega$= 1/2, 5/2 and 7/2 states. More accurate electronic structure computation technique, for example, multi-reference couple cluster (MRCC) method that is considered as the most accurate method to treat the electronic correlation, is also strongly urged to be adopted for this challenge question.

\begin{acknowledgments}
This work is supported by NSFC 61275129, NFSC 21033001, 2012CB821305, NFSC 21203147, and CAS KJZD-EW-W02. Suo and Han would like to thank Professor C. Linton of University of New Brunswick for some beneficial suggestions and bringing new experimental studies of IrO to our attention.
\end{acknowledgments}

\nocite{*}

\begin{thebibliography}{}

\bibitem{Raziunas-JCP-1965}
V. Raziunas, G. Macur, S. Katz, J. Chem. Phys. \textbf{43}, 1010, (1965).

\bibitem{Jansson-JMS-1972}
K. Jansson, R. Scullman, J. Mol. Spectrosc. \textbf{43}, 208, (1972).


\bibitem{Citra-JPCA-1999}
A. Citra, L. Andrews, J. Phys. Chem. A {\bf 103}, 4182, (1999).

\bibitem{Song-TCA-2007}
C. Yao, W. Guan, P. Song, Z. M. Su, Z.J. Feng, L. K. Yan, Z. J. Wu, Theor. Chem. Acc.  {\bf 117}, 115, (2007).

\bibitem{Suo-JPB-2012}
B. Suo, A. Dong, H. Han, Y. Lei, and Y. Wang, Chem. Phys. Lett. {\bf 548}, 12, (2012).

\bibitem{Pang-JPCA-2010}
H. F. Pang, Y. W. Ng, and A. S-C. Cheung, J. Phys. Chem. A {\bf 116}, 9739, (2012).

\bibitem{Adam-JMS-2013}
A. G. Adam, J. A. Daigle, L. M. Esson, A. D. Granger, A. M. Smith, C. Linton, D. W. Tokaryk, J. Mol. Spectrosc. {\bf 286}, 46, (2013).

\bibitem{Suo-supplement}
See discussions in supplement documents.

\bibitem{DKH1974}
M. Douglas, N. M. Kroll, Ann. Phys. {\bf 82}, 89, (1974).

\bibitem{DKH1985}
B. A. Hess, Phys. Rev. A, {\bf 32},756 (1985).

\bibitem{DKH1986}
B. A. Hess, Phys. Rev. A, {\bf 33},3742 (1986).

\bibitem{RASSI}
P.A. Malmqvist, B.O. Roos, B. Schimmelpfennig, Chem. Phys. Lett. {\bf 357}, 230, (2002).

\bibitem{X2C_Liu}
W. Liu, D. Peng, J. Chem. Phys. {\bf 125}, 044102, (2006).

\bibitem{X2C_Ilias}
M. Ilia{\v{s}}, T. Saue, J. Chem. Phys. {\bf 126}, 064102, (2007).

\bibitem{Tyssen-DHF}
J. Thyssen, H. J. Aa. Jensen, Average-of-configurations SCF manuscript, unpublished (1998).

\bibitem{KR-MCSCF}
J. Thyssen, T. Fleig, H. J. Aa. Jensen, J. Chem. Phys. {\bf 129}, 034109, (2008).

\bibitem{KRCI2010}
S. Knecht, H. J. Aa. Jensen, T. Fleig, J. Chem. Phys. {\bf 128}, 014108 (2010).

\bibitem{DIRAC}
{DIRAC}, a relativistic ab initio electronic structure program, Release {DIRAC13} (2013), written by L. Visscher, H. J. Aa. Jensen, R. Bast, T. Saue, with contributions from V. Bakken, K. G. Dyall, S. Dubillard, U. Ekstr{\"o}m, E. Eliav, T. Enevoldsen, E. Fa{\ss}hauer, T. Fleig, O. Fossgaard, A. S. P. Gomes, T. Helgaker, J. K. L{\ae}rdahl, Y. S. Lee, J. Henriksson, M. Ilia{\v{s}}, Ch. R. Jacob, S. Knecht, S. Komorovsk{\'y}, O. Kullie, C. V. Larsen, H. S. Nataraj, P. Norman, G. Olejniczak, J. Olsen, Y. C. Park, J. K. Pedersen, M. Pernpointner, K. Ruud, P. Sa{\l}ek, B. Schimmelpfennig, J. Sikkema, A. J. Thorvaldsen, J. Thyssen, J. van Stralen, S. Villaume, O. Visser, T. Winther, and S. Yamamoto (see http://www.diracprogram.org).

\bibitem{Dunning}
D. E. Woon, T. H. Dunning, J. Chem. Phys.  {\bf 98}, 1358, (1993).

\bibitem{Dyall}
K. G. Dyall, Theor. Chem. Acc. {\bf 112} 403 (2004); K. G. Dyall, A. S. P. Gomes, Theor. Chem. Acc. {\bf 125}, 97, (2009).

\bibitem{CoO1979}
D. W. Green, G. T. Reedy, J. G. Kay, J. Mol. Spectrosc. {\bf 78}, 257, (1979).

\bibitem{CoO1987}
A. G. Adam, Y. Azuma, J. A. Barry, G. Huang, M. P. J. Lyne, A. J. Merer, J. O. Schr\"{o}er, Chem. Phys. {\bf 86},5231, (1987).

\bibitem{CoO1993}
J. Piechota, M. Suffczynski,  Phy. Rev. A, {\bf 48}, 2679, (1993).

\bibitem{CoO1997}
M. Barnes, D. J. Clouthier, P. G. Hajigeorgiou, G. Huang, C. T. Kingston, A. J. Merer,  S. J. Rixon,  J. Mol. Spectrosc. {\bf 186}, 374, (1997).

\bibitem{CoO2005}
S. K. McLamarrah, P. M. Sheridan, L. M. Ziurys, Chem. Phys. Lett. {\bf 414}, 301, (2005).

\bibitem{CoO2006}
J. Guo, T. Wang, Z. Zhang, C. Chen, Y. Chen, J. Mol. Spectrosc. {\bf 240}, 45, (2006).

\bibitem{RhO1998}
X. Li, L.-S. Wang, J. Chem. Phy. {\bf 109}, 5264, (1998).

\bibitem{RhO1999}
A. Citra, L. Andrews, J. Phys. Chem. A, {\bf 103}, 4845, (1999).

\bibitem{RhO2002}
R. F. Heuff, W. J. Balfour, A. G. Adam, J. Mol. Spectrosc, {\bf 216}, 136, (2002).

\bibitem{RhO2003}
R. H. Jensen, S. G. Foug\`{e}re, W. J. Balfour, Chem. Phys. Lett. {\bf 370}, 106, (2003).

\bibitem{RhO2005}
R. F. Heuff, S. G. Foug\`{e}re, W. J. Balfour, J. Mol. Spectrosc. {\bf 231}, 99, (2005).

\bibitem{RhO2007}
J. Gengler, T. Ma, A. G. Adam, T. C. Steimle, J. Chem. Phys. {\bf 126}, 134304, (2007).

\bibitem{RhO2009}
B. Suo, H. Han, Y. Lei, G. Zhai, Y. Wang, Z. Wen, J. Chem. Phys. {\bf 130}, 094304, (2009).

\bibitem{Eugen-JCE-2010}
W. H. E. Schwarz, J. Chem. Educ. {\bf 87}, 444, (2010).

\bibitem{Stefan-Thesis-2013}
S. Knecht, Parallel Relativistic Multiconfiguration Methods: New Powerful Tools for Heavy-Element Electronic-Structure Studies. Dissertation, Mathematisch-Naturwissenschaftliche Fakult\"{a}, Heinrich-Heine-Universit\"{a}t D\"{u}sseldorf, 2009.

\end{thebibliography}



\end{document}


\preprint{JCP/001-IrO}

\title[]{Supplementary materials for ``Relativistic configuration interaction calculation on the ground and excited states of iridium monoxide"}

\author{Bingbing Suo}
\affiliation{Institute of Modern Physics, Northwest University, Xi'an, Shaanxi 710069, China}
\author{Yan-Mei Yu}
\email{ymyu@aphy.iphy.ac.cn,}
\affiliation{Beijing National Laboratory for Condensed Matter Physics, Institute of Physics, Chinese Academy of Sciences, P.O.Box 603, Beijing 100190, China}
\author{HuiXian Han}
\email{hxhan@nwu.edu.cn}
\affiliation{School of Physics, Northwest University, Xi'an, Shaanxi 710069, China}

\date{\today}

\maketitle

%

\section{Compare two lowest states with $\Omega$ value of 5/2 and 7/2 on RASSI level of theory}
Recently, the highly resolution spectra of IrO has been studied by Pang \textit{et al.} (J. Phys. Chem. A, 116, 9379, 2013) and Adam \textit{et al.} (J. Mol. Spectrosc., 286,46,2013). Both of these studies determined that the ground state of IrO had the $\Omega$ value of 5/2. These inspired Suo and Han to reconsider previous calculation (Chem. Phys. Lett., 548, 12, 2012). As pointed out by Adam and coworkers (J. Mol. Spectrosc., 286, 46, 2013), the $^4\Delta_{5/2}$ state may drop below the $^4\Delta_{7/2}$ state when the influence of spin-orbital coupling between the $^4\Delta$ and $^2\Delta$ is included. In previous work by Suo \textit{et. al}, the interaction between the $^4\Delta_{5/2}$ and $^2\Delta_{5/2}$ states is not be considered. Therefore, Suo and Han performed the SOC calculation for $^4\Delta$, $A^2\Delta$ and $B^2\Delta$ states. The results of the $^4\Delta_{7/2}$ and $^4\Delta_{5/2}$ substates are shown in the Table \ref{RASSIRE}. In the new calculation, the energy separation of the 7/2 and 5/2 is reduced to 1108 cm$^{-1}$, which is smaller than 2425 cm$^{-1}$ in previous calculation (Chem. Phys. Lett., 548, 12, 2012). However, we still could not assign a 5/2 ground state due to larger energy gap of the 7/2 and 5/2 states. The discrepancy between theoretical and experimental results may come from insufficient treatment the relativistic effect. The large SOC effect in IrO seems to be calculated by the variational method rather than the perturbation theory. Therefore, multi-configurational electron correlation calculation in the frame of fully relativistic theory is highly desired to give the more accurate description to the ground state of IrO. This is our main motivation of the present work.

\begin{table*}[h]
\caption{Comparing the $^4\Delta_{7/2}$ and $^4\Delta_{5/2}$ substates of IrO.}\label{RASSIRE}
\begin{tabular}{ccccc}
\hline\hline
$\Omega$ & R$_e$ \AA & $\Delta$E & $\omega$ (cm$^{-1}$) & Configurations \\
\hline
7/2 & 1.700 & 0    & 950 & $^4\Delta_{7/2}(100\%)$  \\
5/2 & 1.704 & 1108 & 937 & $^4\Delta_{5/2}(89.6\%),A^2\Delta_{5/2}(9.6\%), B^2\Delta_{5/2}(6.4\%)$ \\
\hline\hline
\end{tabular}
\end{table*}

\section{Excited state of IrO}
More roots are required in the KRCI implementation in order to get the low-lying excited state of IrO, which increases the computation demand greatly. Therefore, the calculation for the excited states is mainly performed on the small basis set, then a composite scheme is adapted to give the estimated value $T^{Com}$ of the excited energy at larger basis set. The correction due to the finite basis set is determined as $\Delta T^{basis}=T^{Ver}_{cv3z}-T^{Ver}_{cv2z}$, where $T^{Ver}_{cv3z}$ and $T^{Ver}_{cv2z}$ are the vertical excitation energies that are calculated at the basis sets [dyall.cv3z (Ir) + aug-cc-pCVTZ (O)] and [dyall.cv2z (Ir) + aug-cc-pCVDZ (O)], respectively. Then, $T^{Com}$ is determined as $T^{Com}=T^{Adi}_{cv2z}+\Delta T^{basis}$, where $T^{Adi}_{cv2z}$ is the adiabatic excited energy obtained at the basis set [dyall.cv2z (Ir) and aug-cc-pCVDZ (O)]. The correction $\Delta T^{basis}$ is regarded as the corresponding uncertainty.

\begin{table*}[btp]
\caption{The excited energy of IrO with $\Omega$=5/2 and 7/2 deduced by the composite technique at the KRCI SD8(13in9)SD level in the MCSCF+CI implementations.}
\begin{tabular}{p{1.2cm} p{1.2cm}  p{1.2cm} p{1.2cm}  p{1.2cm}  p{1.2cm} }\hline\hline
state  &$T^{Adi}_{cv2z}$ &$T^{Ver}_{cv2z}$ &$T^{Ver}_{cv3z}$ & $\Delta T^{basis}$ & $T^{Com}$     \\ \hline
\multicolumn{6}{c}{$\Omega$=5/2}\\
1 	&8481 	&8481 	&8373 	&-108 	&8373  \\
2 	&10350 	&10994 	&12256 	&1262 	&11612 \\
3 	&12691 	&13369 	&14140 	&771 	&13462 \\
4 	&13893 	&14260 	&14752 	&492 	&14385 \\
5 	&14723 	&14901 	&15673 	&772 	&15495 \\
6 	&15448 	&15866 	&16900 	&1035 	&16483 \\
7 	&17231 	&17249 	&17712 	&462 	&17694 \\
8 	&18316 	&18762 	&19750 	&988 	&19304 \\
9 	&20043 	&20519 	&21511 	&992 	&21035 \\
10 	&22022 	&22807 	&23360 	&553 	&22575 \\
11 	&22320 	&23086 	&23699 	&613 	&22933 \\
12 	&22914 	&23735 	&25247 	&1513 	&24426 \\
13 	&23518 	&24470 	&25660 	&1190 	&24708 \\
14 	&24796 	&25935 	&27176 	&1241 	&26037 \\
15 	&25324 	&27778 	&29215 	&1437 	&26761 \\ \hline
\multicolumn{6}{c}{$\Omega$=7/2}\\
1	&8662 	&9165 	&10063 	&898 	&9560  \\
2	&12061 	&12815 	&13880 	&1065 	&13126 \\
3	&15684 	&15684 	&15428 	&-256 	&15428 \\
4	&16209 	&16771 	&17110 	&338 	&16547 \\
5	&17661 	&17963 	&18887 	&924 	&18585 \\
6	&20028 	&21021 	&22006 	&985 	&21013 \\
7	&22319 	&23674 	&24833 	&1159 	&23478 \\
8	&24497 	&24965 	&26014 	&1049 	&25546 \\
9	&26374 	&28992 	&29288 	&296 	&26671 \\ \hline\hline
\end{tabular}
\end{table*}